\documentclass[journal]{IEEEtran}

\usepackage{graphicx} %package to manage images
\usepackage{multirow}
\usepackage{subfigure}
\usepackage{color}
\usepackage{xcolor} %black, blue, brown, cyan, darkgray, gray, green, lightgray, lime, magenta, olive, orange, pink, purple, red, teal, violet, white, yellow
\usepackage{hhline}
\usepackage{url}
\usepackage{amsmath,tabularx}
\begin{document}

\title{A Study of Joint Effect on Denoising Techniques and Visual Cues to Improve Speech Intelligibility in Cochlear Implant Simulation}

\author{Rung-Yu Tseng, Tao-Wei Wang, Szu-Wei Fu, Chia-Ying Lee, and Yu Tsao% <-this % stops a space

\thanks{Rung-Yu Tseng, Tao-Wei Wang, Szu-Wei Fu, and Yu Tsao are with Research Center for Information Technology Innovation, Academia Sinica, Taipei, Taiwan, corresponding e-mail: (yu.tsao@citi.sinica.edu.tw). }
\thanks{Szu-Wei Fu is with Department of Computer Science and Information Engineering, National Taiwan University, Taipei, Taiwan.}
\thanks{Chia-Ying Lee is with Institute of Linguistics, Academia Sinica, Taipei, Taiwan.}% <-this % stops a space  

\thanks{}}

\markboth{}%
{Shell \MakeLowercase{\textit{et al.}}: Bare Demo of IEEEtran.cls for IEEE Journals}

\maketitle

\begin{abstract}
Speech perception is key to verbal communication. For people with hearing loss, the capability to recognize speech is restricted, particularly in a noisy environment or the situations without visual cues, such as lip-reading unavailable via phone call. This study aimed to understand the improvement of vocoded speech intelligibility in cochlear implant (CI) simulation through two potential methods: Speech Enhancement (SE) and Audiovisual Integration. A fully convolutional neural network (FCN) using an intelligibility-oriented objective function was recently proposed and proven to effectively facilitate the speech intelligibility as an advanced denoising SE approach. Furthermore, audiovisual integration is reported to supply better speech comprehension compared to audio-only information. An experiment was designed to test speech intelligibility using tone-vocoded speech in CI simulation with a group of normal-hearing listeners. Experimental results confirmed the effectiveness of the FCN-based denoising SE and audiovisual integration on vocoded speech. Also, it positively recommended that these two methods could become a blended feature in a CI processor to improve the speech intelligibility for CI users under noisy conditions.
\end{abstract}

\begin{IEEEkeywords}
\textit{Speech Intelligibility, Cochlear Implant, Denoising, Speech Enhancement, Audiovisual Integration, Fully Convolutional Neural Network (FCN).}
\end{IEEEkeywords}

% \fi
\IEEEpeerreviewmaketitle

%%%%%%%%%%%%%%%%%%%%%%%%%%%%%%%%%%%%%%%%%%%%%%%%%%%%%%%%%%%%%%%%%%%%%%%%%%%%
\section{Introduction}
\IEEEPARstart
COMMUNICATION is an essential tool that people employ to achieve particular goals, from primary needs to higher-level satisfactions \cite{rubin1998interpersonal}. Verbal communication is among the most efficient means to deliver messages people would like to share. The nature of language acquisition in humans remains fuzzy, but theories from Skinner \cite{skinner1957verbal} and Chomsky \cite{chomsky1965aspects} to relatively modern studies \cite{tomasello2003constructing, ambridge2011child, glenberg2012action} have touched base to depict the speech production in humankind. Using language to communicate, it is easier for people to understand as well as predict others’ actions. Verbal communication, therefore, becomes a crucial process for people to gain social reward in their everyday life \cite{roloff1981interpersonal}. Most importantly, valid verbal communication makes people feel not alone.

Verbal communication relies on two aspects: being able to generate and recognize the speech. Speech perception in humans involves both the internal brain process and external environmental conditions. It had been considered that auditory processing dominates the speech perception until Sumby and Pollack \cite{sumby1954visual} revealed the visual contributions on oral speech intelligibility. In the following decade, the effect of audiovisual integration was tested and proved by Erber \cite{erber1969interaction}. Further, McGurk \cite{mcgurk1976hearing} demonstrated the details on how visual information affects speech recognition. Current research \cite{okada2013fmri, shinozaki2016impact, giordano2017contributions} regarding speech perception has validated that the primary auditory cortex also processes visual information and officially claimed that speech perception was no longer merely hearing. A higher audiovisual gain has been found in speech perception and generation in hearing-impaired children \cite{lachs2001use}. This implies that the brain process of audiovisual integration could be the key for language acquisition and development to support the verbal communication.

The surroundings where people receive sound crucially affects speech perception as well. The speech environment which might cause variability in speech perception is defined by the transmission, such as phones or speakers with accents, and noise conditions \cite{gong1995speech}. Challenges to the clarity of the acoustic speech signals increase the cognitive demands for understanding, and types and levels of background noise are crucial elements causing acoustic difficulties \cite{peelle2018listening}. As a result, studies in speech enhancement (SE) steps in the investigation of speech perception. To improve the recognition accuracy under noisy speech environments, the aim is to minimize the mismatching environmental factors interfering with listeners by enhancing the quality and intelligibility of speech as well as reducing the irrelevant background noise.

Distorted or degraded speech signals were used as the experimental tool in previous studies \cite{foulke1969review, de1982relationship, heiman1986word, dawson2013adaptive} to understand the process of speech recognition in noise . The so-called “unnatural” speech signal plays the role in the system of speech perception to increase the mismatch between acoustic information and the environmental factors, and forces listeners to locate the most reliable components at processing to understand the speech. Meanwhile, distorted speech has shown the level of endurance in human audiences when facing changes in speech structure \cite{remez1981speech, Altmann1993Factors, shannon2007understanding}. The success of research using distorted speech to study speech generation and recognition has been extended from normal hearing (NH) groups to individuals with hearing loss, such as people wearing the cochlear implant (CI) devices\cite{fetterman2002speech, stickney2004cochlear, xu2004tone, wei2007psychophysical, xu2007spectral, chen2015evaluation, mao2017lexical, qi2017relative, ren2018spoken}.

Individuals with hearing loss are limited in their communication, and according to the World Health Organization \cite{world2018addressing}, hearing loss is the fourth highest cause of disability globally. The current estimated population with hearing loss is 466 million worldwide and the number in 2050 is expected to be greater than 900 million if no further action is taken. Under WHO's grades of hearing impairment \cite{olusanya2019hearing}, people with severe-to-profound hearing loss were recommended to wear the hearing aids and CI devices serve as a proved treatment option for them (12 months of age or older) by the Food and Drug Administration (FDA) guidelines. To prevent hearing loss requires controlling risk factors. The Centers for Disease Control and Prevention (CDC) highlights three focus areas: early screening and diagnosis for infants and children, protecting of hearing by recognizing harmful sound levels at home and community, and preventing from occupational noise exposure. From the information provided by the WHO and CDC, hearing loss is an undeniable issue in need of the interference, and noise reduction could be a convincing strategy to slow the growth of hearing loss.

Since speech is a complex representation, speech perception requires higher-level cognitive processing. The feature of noise-vocoding distorted speech \cite{scott2000identification} has allowed researchers destroying the entire intelligibility in the speech to focus on structurally acoustic stimuli. This makes the vocoded materials a promising tool in studying speech perception. Caldwell et al. \cite{caldwell2017assessment} also found that the acoustic challenge caused by spectrally degraded speech could be used to understand the experience of sound quality in CI users. The result could be employed to improve the design of CI devices and further to mitigate relatively poor speech perception for people wearing CI.

The SE process consists of two parts: to enhance the intelligibility and quality of processed speech, and to reduce the noises in the background. Previous well-established algorithms have helped improve the SE in CI users \cite{dawson2011clinical, mauger2012cochlear, chen2015evaluation, goehring2017speech, wang2018supervised, zhao2018deep, goehring2019using, mamun2019convolutional} but there are only few studies with a newly upgrading deep-learning-based algorithm. Traditional SE methods are based on identifying the difference between clean and noisy speech \cite{boll1979suppression, ephraim1984speech, scalart1996speech, rezayee2001adaptive, Xu2015Regression, pandey2019new}. Emerging deep-learning-based models could more accurately match the training and testing conditions to both theoretically and practically optimize the SE performance. For their multi-layered architecture, deep-learning-based models take advantages on extracting representative features to achieve better performance in classification or regression tasks. Speech recognition has become among the typical processes that would benefit from this model \cite{fu2018end}.

The importance to include the visual information in speech perception for CI users could be noticed through the recommendation from the WHO. Other than hearing aids for people with severe-to-profound level of hearing loss, the WHO also advises to at least have lip-reading or signing essential to facilitate the communication \cite{olusanya2019hearing}. Experimental results, further, in children with CI devices have demonstrated to be better multisensory integrators to incorporate visual and audio information at word recognition in speechreading tasks \cite{rouger2007evidence}. The greater audiovisual involvement in speech recognition is expected to advance CI users’ speech perception. Comparing to CI patients, NH participants have exhibited less variation in characteristics of biological, surgical, and device-related elements at performing the tasks \cite{waked2017vocoded}. Assuming similar auditory encoding and processing for both CI and NH groups, the simulated vocoded result by the NH group is able to get nearer the core of cognitive process beyond the unavoidable individual differences.

This study intended to evaluate how speech intelligibility could be improved under denoising SE technology by simulated vocoded corpus on an NH group. A more efficient deep-learning-based denoising algorithm was the main SE process using in current research work. As a pilot study, the experiment was also conducted to test the effectiveness of visual cues in speech perception. In addition, it was anticipated that the cutting-edge deep-learning-based denoising algorithm targets different background noise to help improve human hearing. The task additionally includes two levels of signal-to-noise ratios (SNRs) to understand the threshold of speech perception in noise for listeners. Furthermore, the results from this study could shed light on the possible outcome in the group with hearing loss, particularly for people wearing CI devices.
\par

%%%%%%%%%%%%%%%%%%%%%%%%%%%%%%%%%%%%%%%%%%%%%%%%%%%%%%%%%%%%%%%%%%%%%%%%%%%%
\section{Material \& Methods}

%\graphicspath{ {./images/} }
%\begin{figure}[h]
%  \centering
%     \includegraphics[width=6.6cm]{hddae_new_size.pdf}
%     \caption{Highway-DDAE \ (HDDAE)}
%\end{figure}

Forty participants with gender balance were recruited from the Academia Sinica community to take part in the experiment with the monetary compensation for their time. The group ages
were between 20 and 39 with a mean age of 29.38 years old
(Standard Deviation, SD, =4.63). All participants were native Mandarin speakers with normal or corrected-to-normal vision as well as normal hearing to perceive the stimuli well during the experiment. Except for one left-handed male participant, all others were right-handed. All 40 participants did not report a history of neurological diseases or sensational problems. Written informed consent approved by the Academia Sinica Institutional Review Board for this study was obtained from each participant before conducting the experiment.

The stimuli for this study were video and audio recordings of Mandarin sentences spoken by a native speaker. The source of recordings was based on the Taiwan Mandarin Hearing in Noise Test (Taiwan MHINT, TMHINT, \cite{huang2005development}). All TMHINT materials were unique and each consisted of 10 Chinese characters with the length about three to four seconds. Also, they were specifically designed to have similar phonetic characteristics across the dataset. Among the total of 320 TMHINT utterances, 200 utterances were randomly selected for the SE training set and the remaining 120 utterances for the testing set. The utterances for training and testing sets had no overlap between as well as the types of noise.

The utterances were recorded in a quiet room with sufficient light and the speaker in the video was captured from the front view. Videos were filmed at 30 frames per second (fps) with a resolution of 1920 pixels $\times$ 1080 pixels. Stereo audio channels were recorded at 48 kHz which is precisely the same recording environmental setting as that in Hou et al. \cite{hou2018audio}. The complete experiment included the practice stage and followed by the official testing session with 20 and 100 sentences, respectively. Both selected numbers of utterances for each session were randomly displayed during its part. Each participant was wearing BOSE Triport OE On-Ear Headphone at participating in the experiment.

\graphicspath{ {./images/} }
\begin{figure}[!t]
\centering
\includegraphics[height = 5cm, width = 8cm]{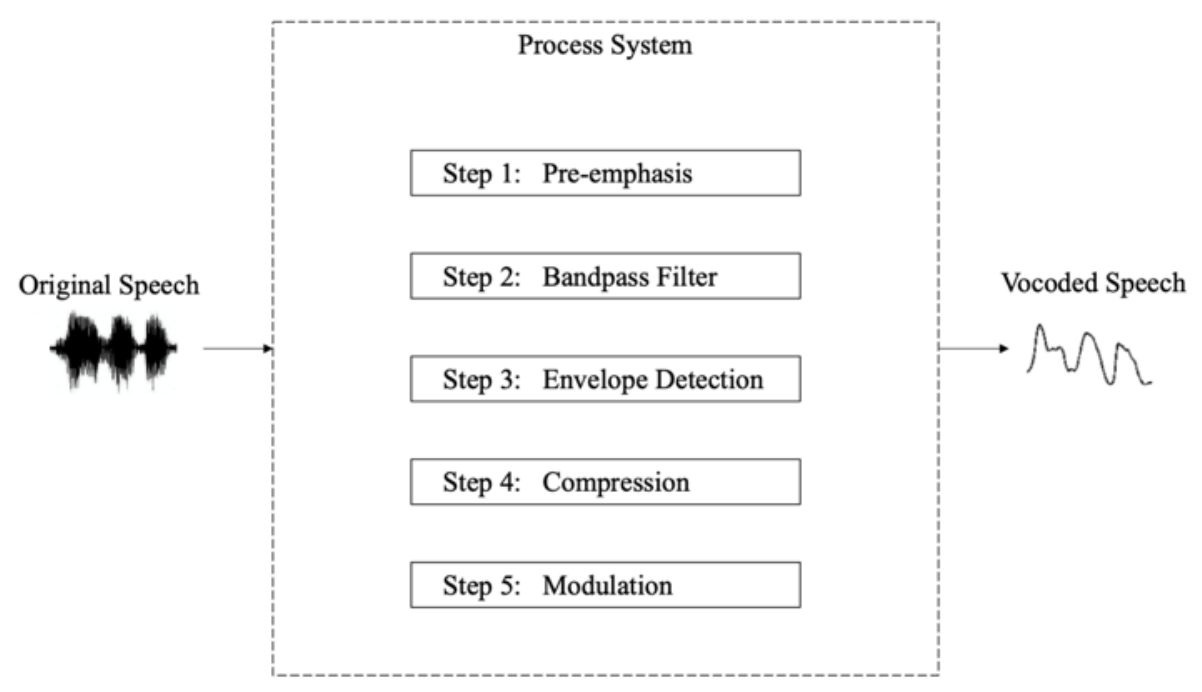} 
\caption{Block diagram of the four-channel tone-vocoder implementation. The first step of speech input is to be processed by the pre-emphasis filter. Then it is filtered by the third-order bandpass filters, and follows to be extracted by a full-wave rectifier. Right after, there comes a second-order lowpass filter, and the strategy ACE is used to compress the envelope of each band. Finally, the modulated compression generates the vocoded speech.}
\label{fig:FIG1}
\end{figure}
\par

The experiment employed a tone-vocoder (Figure 1) as the sound generator to present the stimulus for participants with normal hearing. In the block diagram of a four-channel tone-vocoder, the input signal was first processed through the pre-emphasis filter. Then the  third-order Butterworth bandpass filters filtered the emphasized signal into four frequency bands between 400 and 6000 Hz (with cutoff frequencies of 400, 887, 1750, 3282, and 6000 Hz). The temporal envelope of each spectral channel was extracted by a full-wave rectifier followed by a second-order Butterworth lowpass filter. The envelope of each band was then compressed by the advanced combinational encoder (ACE) strategy in this study.

The ACE strategy continuously varied its compression ratio (CR) on a frame-by-frame basis, with the maximum and minimum values of the compressed amplitude limited within a preset range. The compressed envelopes then modulated the amplitudes of a set of sine waves with frequencies equal to the center frequencies (643, 1319, 2516, and 4641 Hz) of the bandpass filters. Finally, the amplitude-modulated sine waves of the four bands were summed, and the level of the summed signal was adjusted to produce a root-mean-square (RMS) value equal to that of the original input signal.

Using vocoder simulations for the NH listeners under various background noise, speech maskers or numbers of electrodes were found in many previous studies as the proven strategy to understand and further provide basic information on the speech processing in CI users \cite{shannon1995speech, dorman1997speech, dorman1997simulating, fu1998effects, friesen2001speech, stickney2004cochlear, lai2015effects}. However, vocoder simulations were not used for estimating the precise level of performance for each single CI user. This strategy was used to access the performance given particular changing parameters and it allows vocoder simulations to be a valuable tool in CI-related research. Therefore, the tone-vocoder simulation was adapted for NH participants in the current study to understand the possible sound processing in CI users.

To understand the intelligibility of sound processing in a noise environment, three different conditions were used during the listening test: without noise maskers (Clean), with noise maskers (Noisy, masking materials came from the online dataset 'PNL 100 Nonspeech Sounds' \cite{hu2010tandem}), and the SE upon noise maskers. This experiment covered two noise maskers street and engine to represent different noise types (nonstationary and stationary), respectively. The stationary noise means that the whole spectrum of a signal has relatively stable power within any equal interval of frequencies, which is time-independent; while the non-stationary noise owns the opposite characteristics.

\graphicspath{ {./images/} }
\begin{figure}[!t]
\centering
\includegraphics[height = 5cm, width = 8cm]{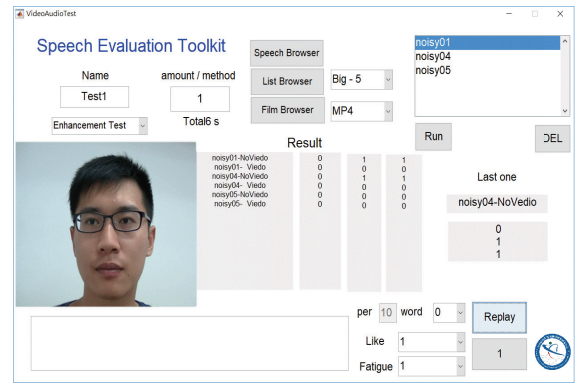} 
\caption{Experimental software. The experimental software is run by using MATLAB. Figure 3 represents the condition with video information. (This experimental toolkit is available via https://github.com/JasonSWFu/VideoAudio)}
\label{fig:FIG2}
\end{figure}
\par

In the experiment, all participants were randomly assigned into two different SNR groups, 1 and 4 dB, with equal numbers of participants in each group. In each SNR group, the testing order for different conditions were put in random for every participants. The 120 TMHNIT utterances in the testing set were prepared with the order of simple randomization into each condition. Test conditions were labelled in all figures throughout this paper as Clean (without any noise masker), FCN\_E (the FCN denoising algorithm targeting the engine noise masker), FCN\_S (the FCN denoising algorithm targeting the street noise masker), Noisy\_E (engine noising masker), and Noisy\_S (street noise masker).

\graphicspath{ {./images/} }
\begin{figure*}[!htbp]
\centering
\includegraphics[height = 6cm, width = 16cm]{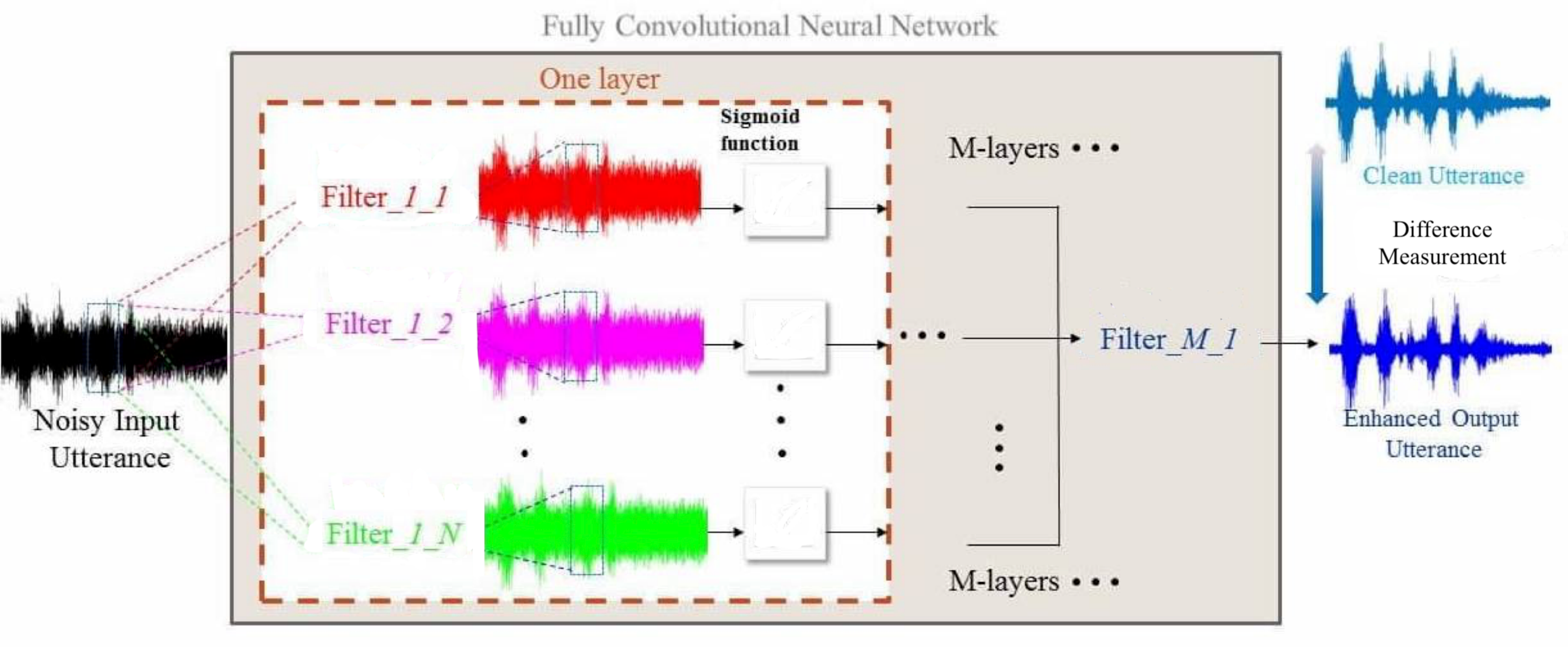} 
\caption{Architecture of utterance-based raw waveform enhancement by the deep-learning-based FCN algorithm. It provides the progress of SE that how noisy input is filtered by multi-layered FCN denoising process.}
\label{fig:FIG3}
\end{figure*}
\par

The interface of the experimental software is shown in Figure 2, and all participants were well instructed to perform the computer-based experiment. During both practice and formal experimental stages, participants were asked to focus on the stimuli by listening to the sound and reading the lips in the video. The stimuli were presented with the sound level of 60 dB, while the level would be adjusted within upper or lower 5 dB upon participants' request. After the stimuli were displayed, participants had to repeat what they recognized accordingly. If they accidentally did not hear or see during the presentation of stimuli, such as clearing the throat or blinking, they would have one more opportunity to re-take it. Once they made the final repetition, the correct answer would be displayed on the screen for checking the accuracy of speech recognition. The accurate character counts were then recorded by choosing the number from zero to ten on the interface.

The fully convolutional neural network (FCN), a deep-learning-based model, was the main algorithm for the SE \cite{fu2018end, gong2019dilated} in this study (the codes for FCN denoising algorithm is available at https://github.com/JasonSWFu/End-to-end-waveform-utterance-enhancement). The FCN was a waveform- and utterance-based denoising SE system. The FCN model could effectively preserve features from local structures with a relatively smaller number of weights for its convolution-layers-only architecture. In addition, FCN convolved the time-domain signal with filters instead of multiplying the frequency representation of a signal by the frequency response of the filter. A feature in the time domain carried much less corresponding energy information than that in the frequency domain, but mainly utilized the relation with its neighbors to represent the frequency concept. This crucial independency stood out that FCN was able to become a more effective denoising algorithm than other conventional fully-connected deep neural networks (DNNs) for waveform-based denoising SE \cite{long2015fully, fu2017raw, fu2018end}.

Most traditional deep-learning models have been designed for a frame-wise process; the result would be less accurate for their problem of incontinuity. The FCN denoising algorithm could fix this by achieving utterance-based enhancement. Furthermore, an FCN could address not merely fixed-length utterances as all fully connected layers were removed in the FCN. This meant, in the FCN denoising algorithm, that input features from different lengths would not have to fit in the matrix multiplication. Assuming that the filter length was \textit{l} and the length of input signal was \textit{L} (without padding), the length of the filtered output would be \textit{L} - \textit{l} + 1. For that FCN contained only convolutional layers, the filters in the operation of the convolution could process inputs with different lengths. More specifically, during the training stage, the FCN-based SE modelcan be trained in an end-to-end utterance-wise manner; while in the testing period, this FCN model can be carried out in a segment-wise form.

In this study, the FCN denoisng algorithm was built to try to incorporate both the mean square error (MSE) and the short-time objective intelligibility (STOI) into the objective function. This aim is to minimize the loss during the training of FCN. The process could be represented by
the equation below:

\begin{equation}
\begin{aligned}
O(w_{u}(t), \hat{w}_{u}(t))&= (\frac{\alpha}{L_{u}} ||w_{u}(t) - \hat{w}_{u}(t)||_{2}^2 
\\ &-stoi(w_{u}(t), \hat{w}_{u}(t) ))
\end{aligned}
\end{equation}

where $w_{u}(t)$ and $\hat{w}_{u}(t)$ are the clean and estimated utterance with index $u$, respectively. $L_u$ is the length of $w_{u}(t)$ (note that each utterance has a different length), and $\alpha$ is the weighting factor of the two targets (which is set the same as used in our previous work \cite{fu2018end}). $stoi(.)$ is the function that calculates the STOI value of the noisy/processed utterance given the clean one. Hence, the weights in FCN could be updated by gradient descent as follows:

\begin{equation}
f_{i,j,k}^{(n+1)}= f_{i,j,k}^{(n)}+\frac{\lambda}{B}  \sum_{u=1}^{B}{ \frac{\partial O(w_{u}(t), \hat{w}_{u}(t) )}{\partial\hat{w}_{u}(t)} \frac{\partial\hat{w}_{u}(t)}{\partial f_{i,j,k}^{(n)}}}
\end{equation}

Where $f_{i,j,k}^{(n+1)}$ is the \textit{i}-th layer, \textit{j}-th filter, \textit{k}-th filter coefficient in
FCN. \textit{n} is the index of the iteration number, \textit{B} is the batch size, and $\lambda$ is the learning rate.

The structure of the overall proposed FCN for utterance-based waveform enhancement was shown in Figure 3, where Filter\_\textit{m}\_\textit{n} denoted the \textit{n}th filter in layer \textit{m}. Each filter coiled together all generated waveforms from the previous layer and then created one further filtered waveform utterance. The goal of SE was to produce one clean utterance, in which the last layer only contained one final filter, Filter\_\textit{M}\_1. This completed end-to-end framework indicated again the efficiency of the FCN denoising algorithm to process the utterance-based enhancement without additional pre- or post-processing.

Compared to other deep-learning-based SE models, FCN provided a better SE result with reduced model sizes \cite{fu2017raw}. Meanwhile, FCN was proven to more effectively enhance speech on non-stationary noises \cite{tsai2017fully}. The STOI was among the major evaluation methods used in related SE studies \cite{taal2011algorithm, gao2016snr}. The STOI measure for FCN indicated a better result than the Noisy condition, particularly for the non-stationary noise type (Figure 4a). In addition, the normalized covariance measure (NCM) was adopted to understand the performance in processing the speech utterances \cite{ma2009objective, chen2010analysis, lai2017deep}. The NCM was based on the covariance between the probing and responsive envelope signals. This metric required only a small number of bands (not limited to a contiguous set) and used simple binary (1 or 0) weighting functions. It was therefore frequently employed to measure the speech intelligibility of vocoded speech \cite{chen2011predicting}. In addition, NCM measure in predicting reliably the intelligibility of noise-suppressed speech was demonstrated its success by Ma et. al. \cite{ma2009objective}. The NCM measure in this study showed consistently higher scores under FCN conditions, especially when targeting a non-stationary noise type (Figure 4b) as STOI did.

\graphicspath{ {./images/} }
\begin{figure}[!t]
\centering
\subfigure[STOI Scores]{
\includegraphics[height = 5cm, width = 8.5cm]
{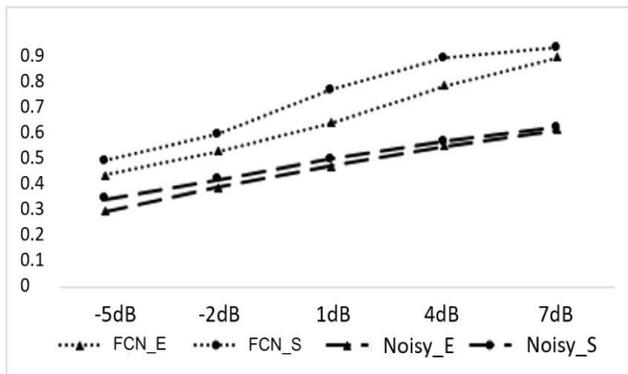}
\label{fig:STOI}
}
\subfigure[NCM scores]{
\includegraphics[height = 5cm, width = 8.5cm] {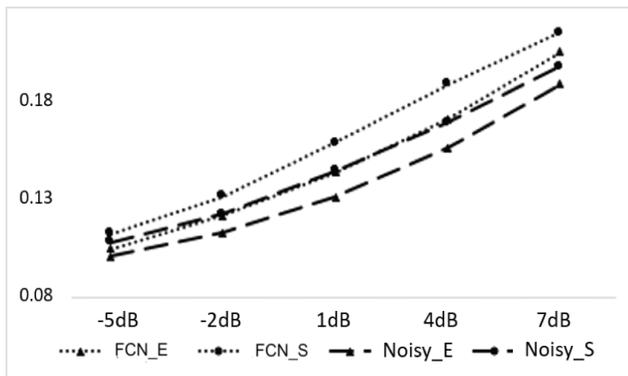}
\label{fig:NCM}
}
\caption{FCN evaluation scores: STOI (a) and NCM (b). Along with the change in SNRs, the FCN received higher scores for both STOI and NCM compared to Noisy conditions, regardless of the types of noise (engine and street); this indicated that the FCN could better facilitate the speech recognition.}
\end{figure}
\par

Not only were the STOI and NCM scores able to quantitatively specify the enhancement resulting from the FCN denoising algorithm in the speech intelligibility, but spectrogram plots and amplitude envelopes also qualitatively showed the advantage of the FCN denoising algorithm. According to Haykin \cite{haykin1995advances}, when studying array processing and signal detection, a time-varying signal could be spectrally represented as a spectrogram. A spectrogram could reveal how noise is reduced to highlight the acoustic characteristics of utterances.

\graphicspath{ {./images/} }
\begin{figure}[!htbp]
\centering
\includegraphics[width=6cm]{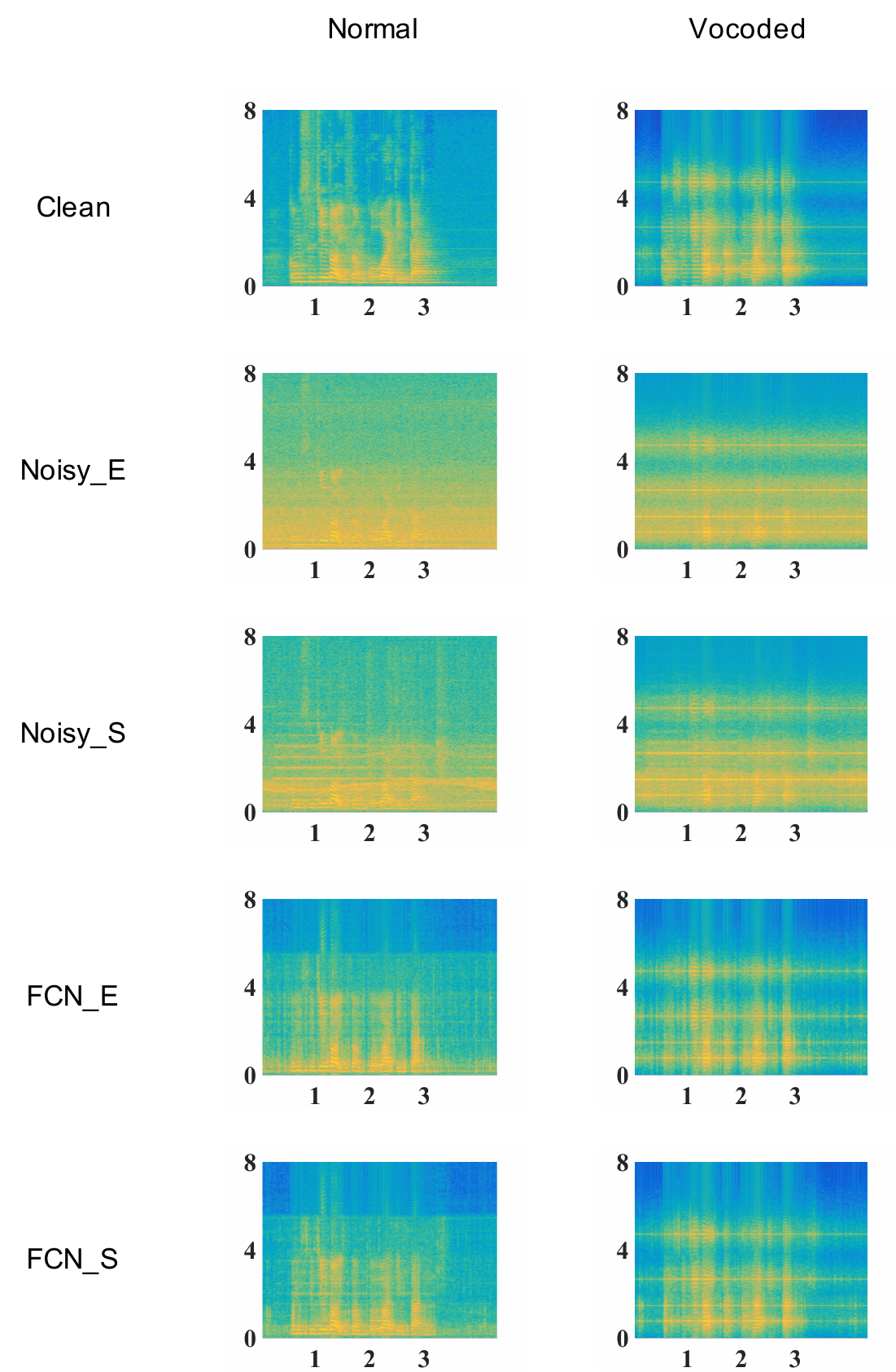}
\caption{Spectrograms of an utterance under different conditions (x axis: time in second, y axis: frequency in kHz). The spectrograms show that FCN denoising algorithm helped reconstruct better utterances under two distinguished types of noise, engine and street, for both original and vocoded speech.}
\vspace{1cm}
\centering
\includegraphics[width=6cm]{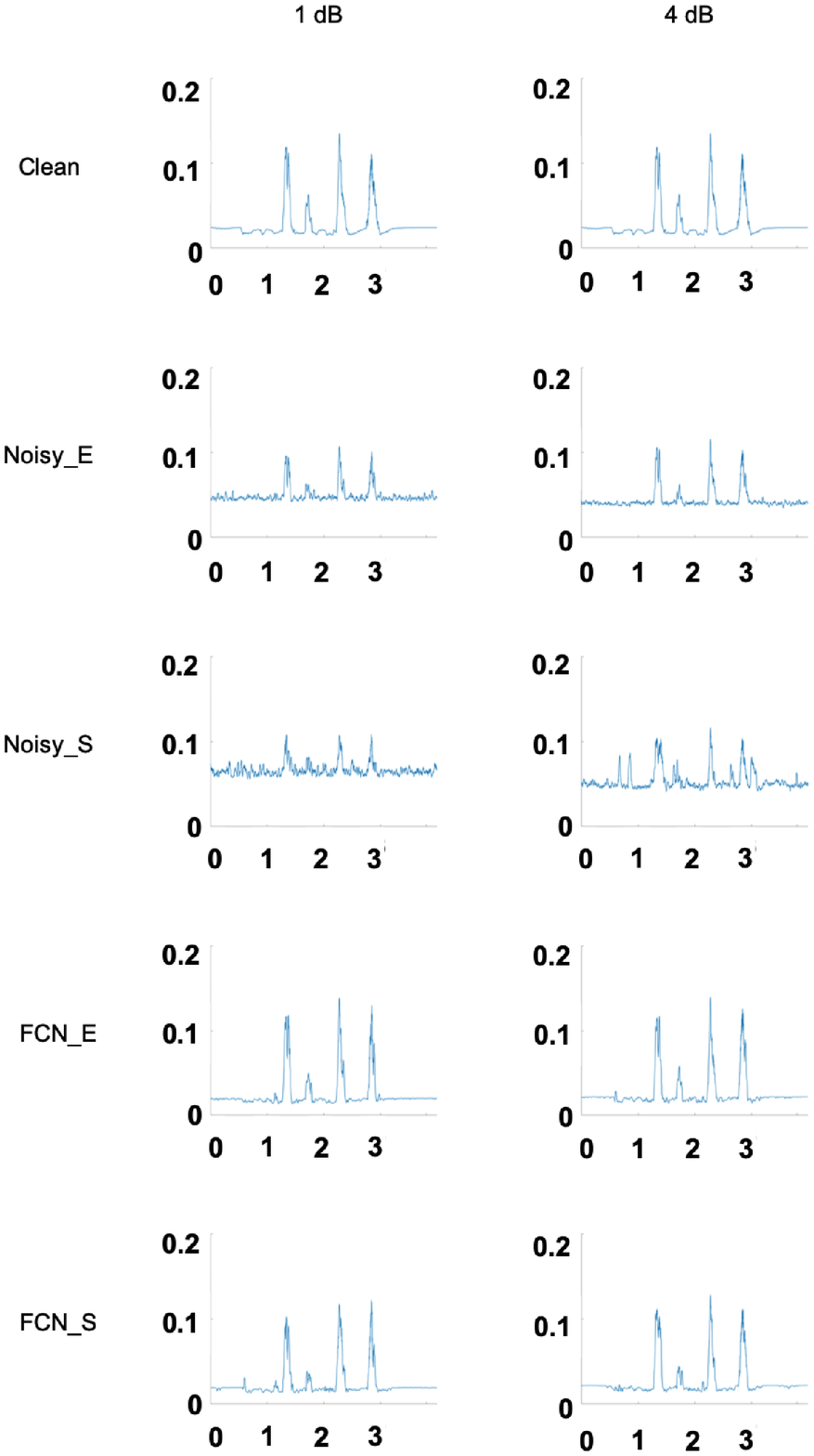}
\caption{Amplitude envelopes from the second-channel frequency band (x axis: time in second, y axis: amplitude). The amplitude envelops from the FCN denoising algorithm resemble to the original clean target sentence, and it indicates the processed speech of better speech intelligibility.}
\end{figure}
\par

As shown in Figure 5, sentences of Clean condition are arranged in the top row of the spectrograms, and other four conditions are followed from the second to bottom rows. With the help of FCN denoising algorithm, the noise maskers were reduced to represent a similar spectral plots as that of the normal utterances (left panel in Figure 5). In addition, the features of each utterance were highlighted under conditions with the FCN denoising algorithm, particularly in vocoded CI (right panel in Figure 5) compared to Normal ones. The spectrogram results demonstrated that the FCN denoising algorithm was able to diminish the noise distortion with less noise residual as shown in the plots. This implied more promising improvement of speech intelligibility via FCN modeling.

From a past research result (American National Standard: Methods for Calculation of the Speech Intelligibility Index \cite{ansi}), the middle-frequency band has a crucial position in the speech intelligibility process. In this study, the four-channel tone-vocoded speech was used to generate sentences as the experimental stimuli to construct a more challenging condition for CI simulation. Given the circumstances, the amplitude envelopes from the second channel were plotted as comparisons for two different SNR tasks, 1 and 4 dBs, under each condition.

The amplitude envelopes provided strong evidence as shown in Figure 6 that after applying FCN denoising algorithm to the target sentence, the waveform was nearer the original shape (a clean condition; the top row for both SNR tasks). The amplitude of each peak was more similar to the source sentence than sentences of the Noisy conditions. In addition, the smaller amplitude could be more clearly represented while it remained distorted under the Noisy conditions. The results of the amplitude envelopes suggest that better speech intelligibility could be achieved using the FCN denoising algorithm.
\par

%%%%%%%%%%%%%%%%%%%%%%%%%%%%%%%%%%%%%%%%%%%%%%%%%%%%%%%%%%%%%%%%%%%%%%
\section{Results}
%%%%%%%%%%%%%%%%%%%%%%%%%%%%%%%%%%%%%%%%%%%%%%%%%%%%%%%%%%%%%%%%%%%%%%

% Fig 1
%\graphicspath{ {./images/} }
%\begin{figure*}[!t]
%  \centering
%     \includegraphics[width=14cm]{MAJDF_projection.pdf}
%     \caption{Online Feature Space Projection \& Back-end Mapping}
%\end{figure*}.  

The descriptive statistic of the listening test showed that the entire performance (mean of 46.91 and SD of 26.34) with the aid of video was better than the audio-only conditions. Participants showed a rather diverse level at conducting different SNR tasks with the separate SD of 25.87 for 1-dB and 25.32 for 4-dB. When working on conditions manipulated as video-aided and audio-only, people’s hearing exhibited a relatively smaller variation in the SD of 19.38 and 20.18, respectively.

\begin{table}[ht]
\caption{\label{tab:table1}ANOVA statistical testing proved that each experimental manipulations (SNR, Video, and Condition) functioned in a statistically significant manner in affecting participants’ performance. In addition, across different conditions, visual aids greatly facilitated to improve the listening test results.}
\begin{center}
\begin{tabular}{lccccccc}
\hline
\hline
&df&Sum Sq&Mean Sq&F value&Pr($>$F)\\
\hline
SNR&1&16154&16154&108.923&$<$2e-16\\
Video&1&121104&121104&816.556&$<$2e-16\\
Condition&4&79604&19901&134.184&$<$2e-16\\
SNR:Video&1&237&237&1.599&0.2068\\
SNR:Condition&4&1006&252&1.696&0.1501\\
Video:Condition&4&1916&479&3.229&0.0126\\
SNR:Video:Condition&4&493&123&0.831&0.5061\\
\hline
\hline
\end{tabular}
\end{center}
\end{table}
\par

The Analysis of Variance (ANOVA, in Table I) indicated that three variables, SNR, Video, and Conditions, used in this study were individually reaching the statistical significance to facilitate the performance of listening test. Notably, the p-value of 0.0126 for Video versus Conditions showed the statistically significant effect on visual facilitation. This confirmed that visual cues were having effect on conditions, but specific aids for each condition were in need of looking into the difference. The paired T-Tests were then conducted to examine the detailed visual aids across different conditions and the effect from the rest of variables, SNR and Conditions, under two SNR groups.

\begin{table}[ht]
\caption{\label{tab:table2}Paired T-Test results for 1 dB. For tasks involving the non-stationary noise type, street, FCN showed better facilitation compared to Noisy conditions, particularly when there were no visual cues to further help people’s hearing.}
\begin{tabular}{cccccccc}
\hline
\hline
Video&Noise&Condition&Mean&SD&t&df&p-value\\
\hline
\multirow{4}{*}{Yes} & \multirow{2}{*}{Street} & FCN & 51.90 & 16.28 & \multirow{2}{*}{2.3664} & 19 & \multirow{2}{*}{0.02874} \\
& & Noisy & 44.60 & 12.98 & & 19 & \\
& \multirow{2}{*}{Engine} & FCN & 53.95 & 18.02 & \multirow{2}{*}{1.2257} & 19 & \multirow{2}{*}{0.2353}\\
& & Noisy & 50.60 & 16.38 & & 19 & \\
\multirow{4}{*}{No} & \multirow{2}{*}{Street} & FCN & 18.50 & 9.74 & \multirow{2}{*}{3.8595} & 19 & \multirow{2}{*}{0.001056} \\
& & Noisy & 11.00 & 6.18 & & 19 & \\
& \multirow{2}{*}{Engine} & FCN & 17.55 & 7.92 & \multirow{2}{*}{0.64491} & 19 & \multirow{2}{*}{0.5267}\\
& & Noisy & 16.10 & 8.46 & & 19 & \\
\hline
\hline
\end{tabular}
\end{table}
\par

The paired T-Test results provided more information regarding how the FCN denoising algorithm facilitates human’s listening performance compared to the Noisy condition. In Table II, the T-Test results for the lower-SNR tasks indicated that particularly under the non-stationary noise type, street, the FCN better helped people during the listening test. Regardless of visual aids, conditions of FCN targeting street noise both reached the statistical significance with p-values of 0.02874 (with video) and 0.001056 (without video), respectively. It was noteworthy that when lacking visual aids for people’s hearing, the benefit from FCN became more recognizable as the p-value reached its most practical statistical significance ($<$ 0.01). The result of higher-SNR tasks was listed on the Table III. It was similar to the lower-SNR results, yet only one statistical significance, as p-value of 0.02611, had been reported on the condition FCN targeting street noise when there was no visual aids.

\begin{table}[ht]
\caption{\label{tab:table3}Paired T-Test results for 4 dB. The overall tendency had similar results as 1 dB but with higher p-values. This confirmed again that FCN functioned as a reliable facilitator, especially when having background noise such as the non-stationary noise type, street, and the lack of other aids such as visual cues.}
\begin{tabular}{cccccccc}
\hline
\hline
Video&Noise&Condition&Mean&SD&t&df&p-value\\
\hline
\multirow{4}{*}{Yes} & \multirow{2}{*}{Street} & FCN & 70.10 & 12.68 & \multirow{2}{*}{1.7302} & 19 & \multirow{2}{*}{0.09981} \\
& & Noisy & 64.50 & 12.39 & & 19 & \\
& \multirow{2}{*}{Engine} & FCN & 64.10 & 12.29 & \multirow{2}{*}{-1.4236} & 19 & \multirow{2}{*}{0.1708}\\
& & Noisy & 67.65 & 12.36 & & 19 & \\
\multirow{4}{*}{No} & \multirow{2}{*}{Street} & FCN & 29.70 & 9.71 & \multirow{2}{*}{2.4127} & 19 & \multirow{2}{*}{0.02611} \\
& & Noisy & 24.25 & 11.75 & & 19 & \\
& \multirow{2}{*}{Engine} & FCN & 26.70 & 9.99 & \multirow{2}{*}{-0.33007} & 19 & \multirow{2}{*}{0.745}\\
& & Noisy & 27.70 & 13.12 & & 19 & \\
\hline
\hline
\end{tabular}
\end{table}
\par

The higher- and lower-SNR tasks had resembled results in the paired T-Test. It was only that greater difference between the FCN and Noisy conditions under the non-stationary noise type, street, to be proved statistically significant for the 1-dB task regardless of visual aids and 4-dB task without video. However, the performance of FCN targeting engine noise was not differentiated from the Noisy condition statistically.

The insignificant T-Test results for these two distinct SNR tasks suggested that there might be some preference for FCN targeting engine noise. When it came without the visual aid, the p-values of FCN targeting engine noise were 0.745 for 4-dB versus 0.5267 for 1-dB tasks. Another set of outcome for the tasks with the visual aid were 0.1708 for 4-dB versus 0.2353 for 1-dB tasks. In statistics, lower p-value means stronger evidence in favor of alternative hypothesis to imply the effect of experimental manipulation. In conditions with video aids, lower p-value reasonably appeared in 4-dB tasks. For the conditions without video aids, however, lower p-value fell in the 1-dB tasks to hint that FCN better helped the hearing during lower-SNR even with no additional visual information. For the condition of FCN targeting engine noise, the variation of p-values in different SNR groups might reveal a tendency about the degree of improvement by FCN.

The percentage of accuracy in listening test was drawn in Figure 7, and the provided results were alike as for the statistical testing. According to the Figure 7a, results of lower-SNR tasks showed generally greater scores in the FCN compared to the Noisy conditions, regardless of the types of noise masker. However, taking the visual information into consideration, participants was doing better in FCN targeting engine noise; while without visual aids, the higher accuracy fell in the condition of FCN targeting street noise. The performance for higher-SNR tasks unveiled another tendency in Figure 7b. The highest accuracy still occurred at the condition FCN targeting the non-stationary noise masker, street. However, the results here showed that not both FCN conditions dominated in the 4-dB tasks as them did in the lower-SNR tasks.

\graphicspath{ {./images/} }
\begin{figure}[!htbp]
\centering
\subfigure[1 dB]{
\includegraphics[height = 5cm, width = 8.5cm]
{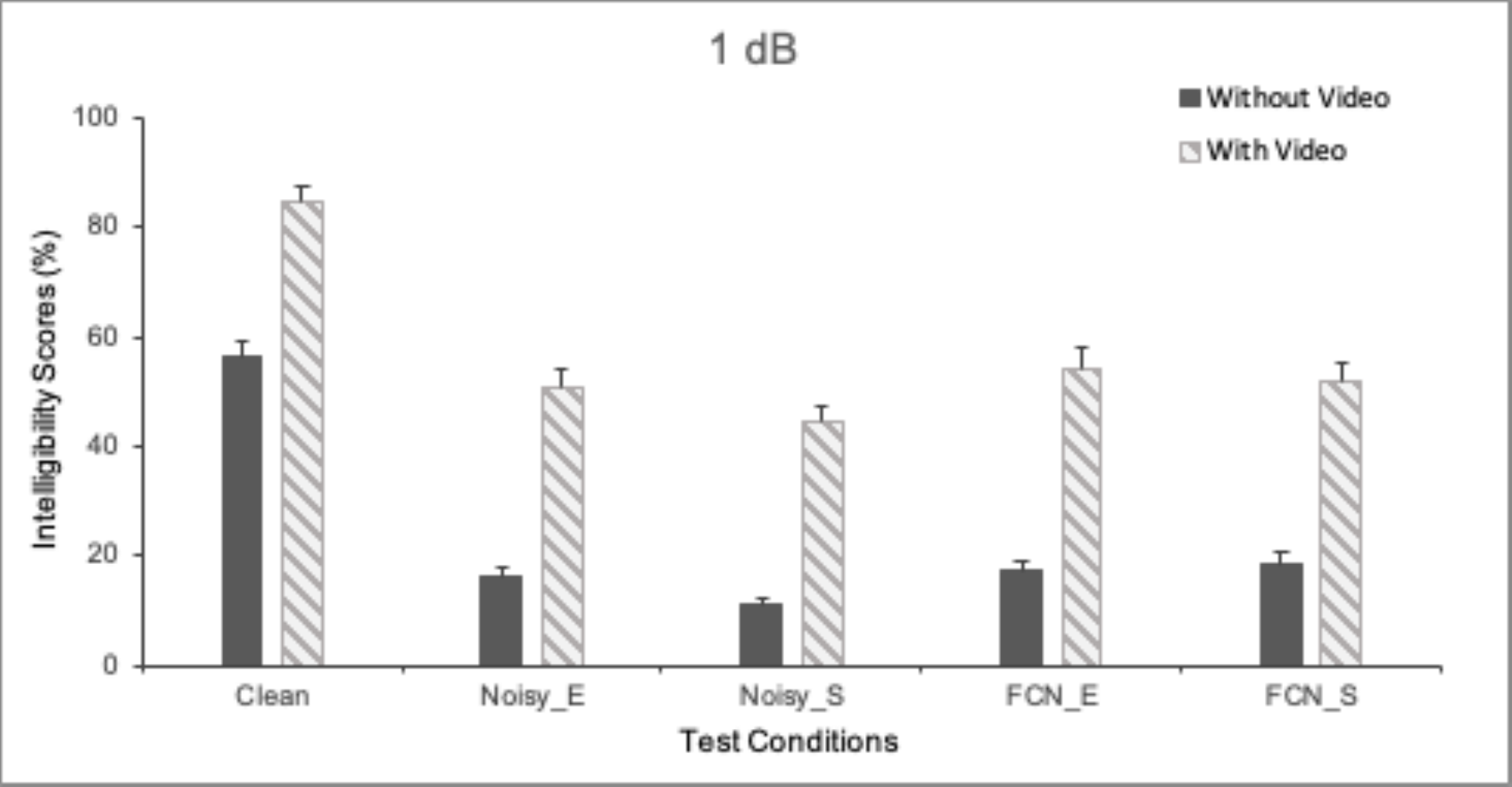}
\label{fig:1_dB}
}
\subfigure[4 dB]{
\includegraphics[height = 5cm, width = 8.5cm] {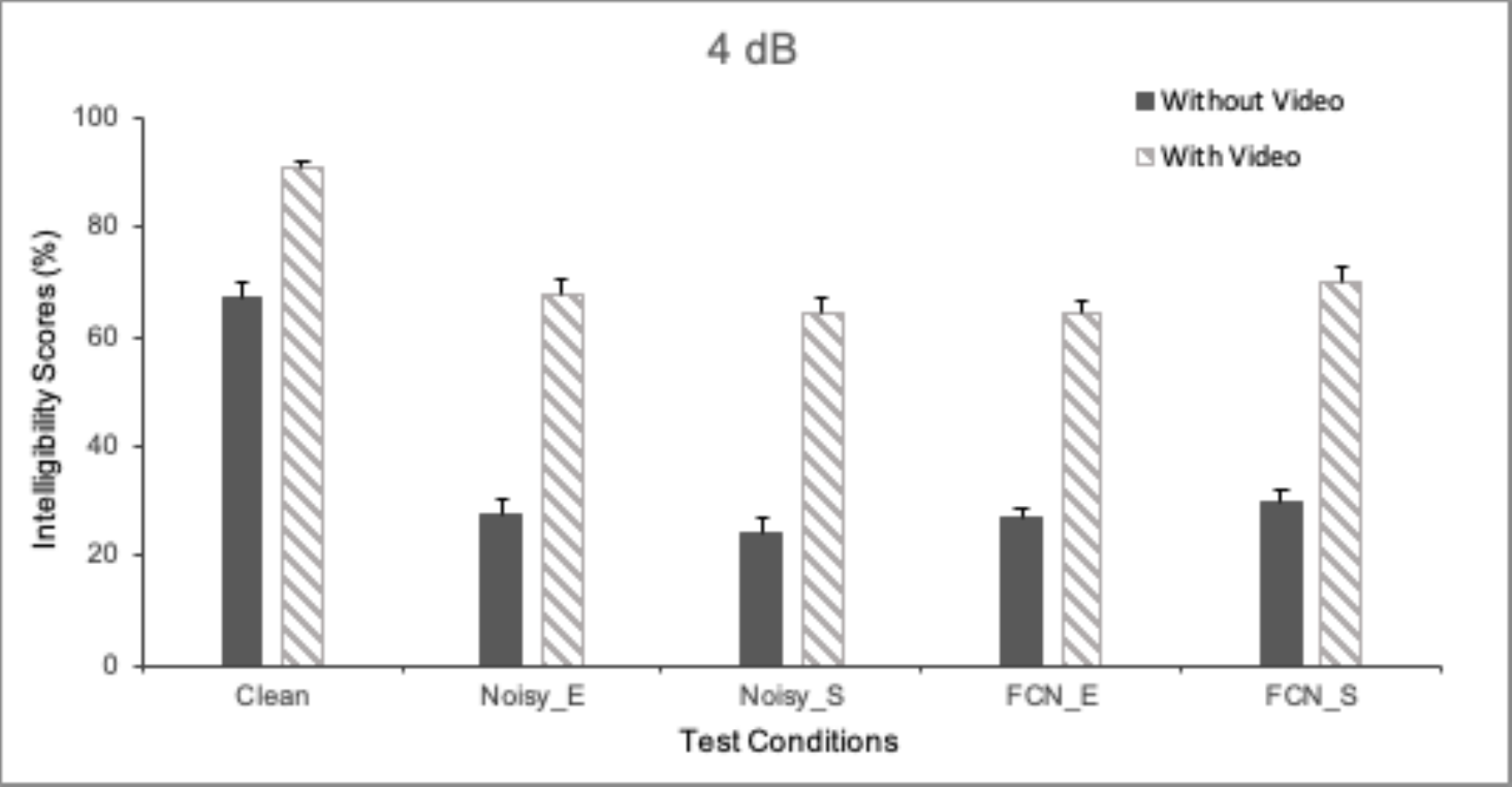}
\label{fig:4_dB}
}
\caption{Listening test results: performance of (a) 1-dB tasks and (b) 4-dB tasks. Conditions with video are marked in a diagonal pattern for both SNRs while solidly filled bars indicate conditions without visual information.}
\end{figure}
\par

The analysis of the ANOVA and the paired T-Test presented no statistically significant indication for the FCN targeting the stationary noise type, engine, but the improved performance through the facilitation by FCN did exist. The interaction plot was to show that two independent variables interact if the effect of one of the variables varies depending on the level of the other variable. Figure 8 displayed the interaction between variables Condition and SNR. During the higher-SNR task, which were relatively clear for human hearing, the FCN targeting engine noise was merely in the third spot of all the effective conditions. As tasks moved toward lower-SNR, however, the FCN targeting engine noise became a better performer while the other three conditions remained the same ranking. This matched the paired T-Test results that there might be a possible preference for FCN targeting engine noise to step in the facilitation of human hearing.

\graphicspath{ {./images/} }
\begin{figure}[!t]
\centering
\includegraphics[height = 5cm, width = 8cm]{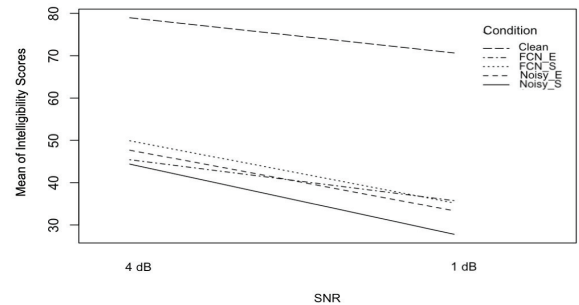} 
\caption{Interaction plot between two distinct SNRs over different conditions. FCN targeting engine noise better facilitated human hearing during the lower-SNR tasks while the other three conditions remained the same ranking in spite of the change of different SNR tasks.}
\label{fig:FIG8}
\end{figure}
\par

Both inferential and descriptive statistics in different SNR tasks suggested the ease of a higher-SNR sound to be caught by people’s hearing. In general, for both SNRs, the Clean condition without any noise masking took participants the least effort to hear the sounds; however, visual information helping improve hearing was overwhelmingly across every single condition, no matter which SNR tasks were involved.

The listening test scores and statistical results of ANOVA and the paired T-Test all demonstrated that the FCN was able to serve as an effective denoising algorithm and helped enhance the intelligibility of speech recognition. Furthermore, the paired T-Test results and the interaction plot provided more clues regarding how the FCN contributed to human hearing and what conditions might be the best fit for FCN involvement.
\par

%%%%%%%%%%%%%%%%%%%%%%%%%%%%%%%%%%%%%%%%%%%%%%%%%%%%%%%%%%%%%%%%%%%%%%%%%%%%
\section{Discussion}

Consistent with past research \cite{chen2008seeing, desai2008auditory, tremblay2010audiovisual}, visual information is of great help in facilitating people’s hearing. In current listening test results, the performance was improved with the aid of visual cues across various conditions. However, the level of facilitation differs. First, the visual information works well even as background noise appeared and helps particularly better in tasks with specific types of noise maskers. For higher-SNR tasks, with the help of visual information, listeners are able to considerably improve their performance under the non-stationary noise type, street, with the FCN denoising process. The effectiveness of visual cues shows the most extent compared to other conditions.

The performance of both lower- and higher-SNR tasks reveals that there might be a critical threshold for listeners to detect the sound in noise. In the result of lower-SNR tasks, the support from the FCN denoising is manifest for both types of noise maskers. The possible reason could be that 1 dB is too challenging for listeners to differentiate the background noise from the targeting sounds; both background noise and targeting sounds become homogeneous during sound processing. Visual cues and FCN help sharpen the targeting sounds for listeners to distinguish them from the background noise. Alternatively, the higher SNR task allows participants to rather effortlessly hear both the sound and noise; therefore, the boundary of the noise and targeting sound emerges. People with normal hearing can more easily process the target-perceiving and denoising in higher-SNR tasks. The threshold for the NH and CI groups might not be the same, but the critical threshold is an important clue to enhance the speech intelligibility.

This study also collected evidence from the post-analysis to reconsider the function of the FCN denoising algorithm for different SNR tasks. The listening test scores and the interaction plot revealed the FCN targeting engine noise was slightly higher than the FCN targeting street noise within lower-SNR tasks. The result implied that in spite of the effect of the FCN targeting engine noise was not universally observed across different conditions, the weighting of FCN effectiveness could become more obvious once people have less cues or more interrupting background noise for them to understand the targeting sounds. As a result, the listening test performance indicates that the FCN denoising algorithm works differently toward alternative types of noise in higher-SNR tasks but dominates in lower-SNR tasks as participants need the enhancement to detect the comparatively weaker line between targeting sounds and background noises as the phenomenon of stochastic resonance \cite{gammaitoni1998stochastic, zeng2000human, behnam2003noise}. That is, the FCN denoising algorithm works particularly competent in a noisy listening environment.

The FCN denoising algorithm plays a role to potentially improve participants’ performance in the listening test. Given the noise interference, participants’ performance under FCN conditions was the best among the test results, for both lower- or higher-SNR tasks. In addition, the accuracy rate of the FCN conditions is generally higher when involving background noise such as street sounds, a non-stationary noise type. This matches the results of Tsai \cite{tsai2017fully}'s previous study that the FCN extracts cleaner speech to achieve an improved listening test result, particularly for a non-stationary noise type. Comparing to purely Noisy conditions, listeners hear better under the stationary noise type. The listening test results provide more confidence to record the effectiveness of the FCN denoising algorithm in enhancing the speech perception.
\par

%%%%%%%%%%%%%%%%%%%%%%%%%%%%%%%%%%%%%%%%%%%%%%%%%%%%%%%%%%%%%%%%%%%%%%%%%%%%
\section{Conclusions}

The FCN algorithm is demonstrated as a better denoising SE model as it is similar to a traditional CNN but not limited to process fixed-length inputs \cite{fu2018end}. Given the flexibility that FCN can contribute, the denoising technology has been leveled up and the listening test results in this study further prove the effectiveness of FCN in vocoded speech intelligibility. In addition, under specified noise maskers, conditions with FCN were able to provide listeners more enhanced speech perception to obtain higher accuracy in scores. Since the preliminary result in CI simulation is positive in verifying the superiority of the FCN, having CI users participate in the future investigation is the most empirical means to determine the real effect on the group with hearing loss.

The future implement based on the results from this paper will be anticipating in two levels. First step is to transform the finding about the joint effect onto audiovisual SE. Hou et al. \cite{hou2018audio} have reported an audiovisual SE model using CNN to generate enhanced speech and then reconstruct to images. In current study, the SE model was based on FCN and the following work should be to build the fused encoder-decoder FCN-based SE model for audiovisual SE. After completing the audiovisual SE, during the next stage, it should be to apply the ready audiovisual integration algorithm onto multi-devices for both ears and eyes.

As all might be aware that most CI users have only hearing impairment with no dysfunctions for other senses, such as sight, smell or touch, and it seems reasonable that wearing aids for users are mainly to facilitate their hearing. However, as our brain process the information in the way of sensory integration, the vision of CI users is used to help their weakened auditory sense. In addition, during the situations that CI users receive less or none visual information, for instance, conversations during the phone call, it is necessary for them to have the help from multi-devices. CI users are able to expect better hearing from both the FCN denoising algorithm within their CI processor and converted images to show enhanced speech visually on their wearing goggles. This study expects to further provide the scientific evidence to include the visual information in the assistive auditory devices with better help for CI users.

As a pilot study for audiovisual-aided vocoded speech intelligibility, the listening test results have successfully demonstrated its improvement. The performance in the experiment validates the power of audiovisual integration to enhance vocoded speech intelligibility by the much better accuracy rate under conditions with visual aid. This experimental evidence strongly implies the possibility of an updated audiovisual CI system. Applying the human process of audiovisual information onto current operating algorithm in CI processors is a means to better the speech perception for people wearing CI devices. In addition, given this encouraging listening test results in CI simulation, one can envision an advanced fusion system joining deep-learning-based FCN denoising algorithm and audiovisual integration to further boost the speech intelligibility for the hearing-loss group.

To consolidate the updated fusion system, functional optimization of the FCN denoising modeling is a necessary future work as current CI devices remain multiple engineering issues on the speech processor \cite{zeng2004trends, zeng2008cochlear}. The deep structure, however, still requires more computational hardware needs and higher costs than those of traditional models. To properly allocate these resources, developing quantization techniques were used to compress the model \cite{hsu2018study}. With the help of quantized deep-learning-based model, speech intelligibility would display a more progressive improvement in its reduced processing time \cite{ko2017precision}. The fusion of audiovisual integration and denoising SE modeling could be working competently in CI devices as the revamped hardware is evolved progressively in the near future.

Speech intelligibility is crucial for both NH and hearing-loss groups to manage the conversations in social interaction, and denoising technology serves as a tool to improve the interpersonal communication by enhancing the quality of hearing. Beneficial from the development of a deep-learning technique, the FCN denoising algorithm makes progress based on the advantages of conventional modelings to advance towards a more promising enhancement for speech intelligibility. In addition, the impact of visual cues on enhancing the vocoded speech is clearly proven through the experiment. The listening test results in this CI simulation provide solid evidence that both audiovisual integration and SE technology could greatly facilitate people’s hearing even in a noisy environment. To the final goal to contribute to a hearing-loss group, applying both audiovisual information and FCN in a future investigation involving CI users is a foreseeable process to determine the true value of deep-learning-based modeling for SE and the influence of audiovisual integration.
\par

\section*{Acknowledgment}
This work was supported in part by the Ministry of Science and Technology of Taiwan under Grant MOST 106-2221-E-001-017-MY2,  107-2221-E-001-012-MY2, 108-2628-E-001-002-MY3, and 108-2811-E-001-501-. The authors would also like to sincerely thank Dr. Kevin Chun-Hsien Hsu from the Institute of Cognitive Neuroscience, National Central University, and Dr. Hao Ho from the Institute of Statistical Science, Academia Sinica, for providing insight and expertise that greatly assisted this research.


%\ifCLASSOPTIONcaptionsoff
%  \newpage
%\fi



%\end{IEEEbiography}


%\newpage



%\vfill



\bibliographystyle{IEEEbib}
%\bibliography{ref}

\end{document}